\documentclass[letterpaper]{mc2025}

\usepackage{algorithm}
\usepackage{algorithmic}

\title{Randomized Quasi-Monte Carlo Sampling in The Random Ray Method for Neutron Transport Simulations}

\addAuthor[sam.pasmann@thea.energy]{Samuel Pasmann}{1}
\addAuthor{John Tramm}{2}
\addAffiliation{1}{Thea Energy, Kearny, NJ}
\addAffiliation{2}{Argonne National Laboratory, Lemont, IL}

\Abstract{%
In a random ray method of neutral particle transport simulation, each iteration begins by sampling a set of rays before proceeding to solve the characteristic transport equation along the linear paths the rays follow. Historically, traditional pseudorandom number generators have always been used when sampling ray starting points and directions. In the present work, we experiment with using a randomized Quasi-Monte Carlo sampling technique in place of the traditionally fully random technique with the goal of reducing the variance of simulation results. To evaluate the numerical and runtime performance of the new sampling scheme, it is implemented in the random ray solver of the OpenMC particle transport application and tested on the 2D C5G7 benchmark simulation. The new scheme delivered over a 10\% reduction in the average pin-cell error and up to a 8\% overall speedup compared to the state-of-the-art, owing both to numerical reductions in uncertainty as well as practical reductions in shared memory contention.
}

\keywords{Neutron Transport, The Random Ray Method, Quasi-Monte Carlo}

\shortTitle{RQMC Sampling in The Random Ray Method}

\authorHead{S. Pasmann \& J. Tramm}

\begin{document}

%%%%%%%%%%%%%%%%%%%%%%%%%%%%%%%%%%%%%%%%%%%%%%%%%%%%%%
\section{Introduction}\label{sec:introduction}

\subsection{The Random Ray Method}

The random ray method (TRRM) is a recently developed method for neutral particle transport~\cite{TRRM}. It is a stochastic multigroup method closely related to the Method of Characteristics (MOC)~\cite{askew}. Similar to MOC, the random ray method decomposes the governing Boltzmann partial differential equation into a family of ordinary differential equations that each apply along only a single line. In the method of characteristics, a deterministic quadrature is used to select which characteristic lines to solve for, with the same quadrature being reused for every power iteration in the simulation. The random ray method instead uses a stochastic quadrature that changes each iteration, making it a fully stochastic method requiring inactive and active cycles in a manner similar to a Monte Carlo simulation. More information on the random ray method, including its mathematical derivation, can be found in~\cite{TRRM}. The random ray method has been implemented in the ARRC~\cite{ARRC}, SCONE~\cite{scone_openmc_random_ray}, MPACT~\cite{mpact}, and OpenMC~\cite{scone_openmc_random_ray} particle transport applications.

The random ray method has been a topic of research in recent years as it has recently been proven to be a highly numerically efficient method as compared to existing deterministic methods~\cite{TRRM,ARRC,scone_openmc_random_ray}. This surprising advantage stems from random ray being able to run with an extremely coarse ray density as compared to the much finer requirements of a deterministic simulation. This can be accomplished while maintaining accuracy due to the fact that the random quadrature changes each iteration yet still improves the iterative estimate of the source term.

To date, the random ray method has always been implemented using a fully stochastic pseudo-random number generator for sampling ray starting locations (which are uniformly sampled in space and angle throughout the simulation domain). Once sampled, the rays traverse the simulation domain along a linear path, solving the MOC equation as they pass through each cell, until terminating after a user specified distance. In the present study, we will implement an alternative sampling technique for generating ray starting distributions in the random ray solver of OpenMC, and evaluate the impact that the new technique has on numerical and runtime performance.
 
\subsection{Quasi-Monte Carlo Methods}
Quasi-Monte Carlo (QMC) techniques use low-discrepancy sequences in place of typical pseudo-random number generators in Monte Carlo techniques. Like pseudo-random number generators, low-discrepancy sequences are intended to produce $N$ points $U[0,1]^D$. Unlike pseudo-random numbers, low-discrepancy sequences use deterministic algorithms designed to maximize the distance between the points which is measured using discrepancy statistics~\cite{chen2014panorama}. The decreased discrepancy of the sample set results in variance reduction in QMC integration approximations and in ideal problems can converge the sampling error close to a rate of $O(N^{-1})$, an improvement over the $O(N^{-1/2})$ expected from pseudo-random samples in Monte Carlo techniques~\cite{caflisch1998monte}.

Various low-discrepancy sequences have been developed each with unique characteristics. Common sequences include the Halton sequence~\cite{halton1960efficiency}, Sobol sequence~\cite{sobol1967distribution}, and Latin hyper-cube~\cite{mckay2000comparison}. Some sequences like the Sobol and Halton, are extensible in both $N$ and $D$, however the equidistribution qualities of both sequences begin to degrade in higher dimensions (typically starting around $D>10$). Fortunately, this is less of a concern in ray-tracing applications which typically require a five-dimensional sequence (three in space and two in angle). 

Despite a significantly improved convergence rate, QMC techniques are not commonly used in particle transport applications~\cite{Spanier1995}. This is because each point generated in the sequence is dependent on the previous and it's inadvisable in QMC to split the sequence and use separate chunks as individual sample sets. This makes QMC techniques ill-suited for modeling the particle random-walk process when scattering is present. QMC techniques must be implemented in applications that are not Markovian processes or steps must be taken to ensure the Markovian assumption is maintained. Recently, the iterative-Quasi Monte Carlo method, a hybrid-method for multigroup neutron transport simulations, employed QMC sampling in a similar ray-trace procedure to TRRM. iQMC can converge the scalar flux solution at the expected $O(N^{-1})$, but must also use a fixed-seed approach in the transport sweep due to the deterministic nature of the low-discrepancy sequence~\cite{Pasmann2023Quasi}.

\subsection{Randomized Quasi-Monte Carlo Methods}
TRRM requires that rays are sampled at different locations and traveling in different directions each iteration. This requirement combined with the deterministic nature of QMC and its poor performance in higher dimensions, limits its direct applicability to TRRM. Instead, this study utilizes randomized-QMC (RQMC) techniques. RQMC techniques have seen increasing interest and research over the last few decades~\cite{lecuyer2018Randomized} and are designed to construct each point individually $U\left[0,1\right]^D$ while collectively the $N$ points retain their low discrepancy. By randomizing the sequence each iteration, we can generate a unique set of RQMC samples for each set of particles. Additionally, by randomizing the QMC samples we can estimate the variance of the integral, something that is difficult to do with unrandomized QMC.

Somewhat counter-intuitively, RQMC techniques have been shown to obtain a root mean square error lower than that of standard QMC techniques. This is generally explained by idea that random permutations of the samples cancel some of the systematic errors introduced from the low-discrepancy sequence algorithm~\cite{owen2021strong}. However, RQMC techniques do not escape the curse of dimensionality and show degraded performance in higher dimensions~\cite{dimov2008monte}. While RQMC samples should still not be used to model Markovian processes like a particle random-walk, RQMC techniques were recently applied in the iQMC method to great benefit~\cite{pasmann2024development}.

\begin{figure}[ht!]
  \centering
  \includegraphics[width=0.9\textwidth]{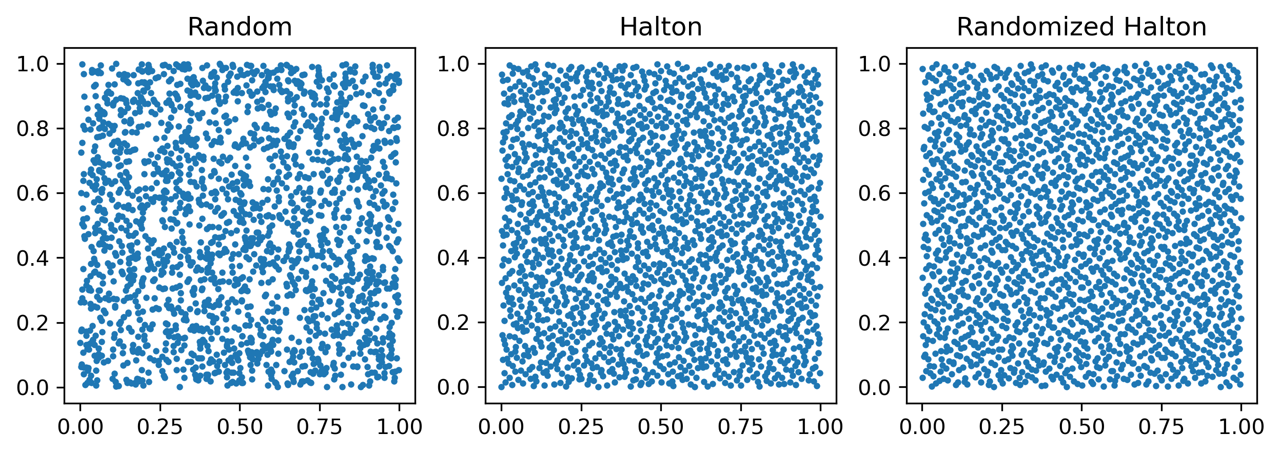}
  \caption{Points generated in the unite square from a pseudo-random number generator, Halton Sequence, and randomized Halton sequence.}
  \label{fig:samples}
\end{figure}

%%%%%%%%%%%%%%%%%%%%%%%%%%%%%%%%%%%%%%%%%%%%%%%%%%%%%%
\section{METHODOLOGY}\label{sec:methodology}
Several RQMC methods have been developed to compliment traditional QMC techniques~\cite{lecuyer2002recent}. This study utilizes Owen's randomization method of the Halton Sequence~\cite{Owen2017}. Owen’s randomization introduces independent random permutations to the samples, making it possible to generate $R$ distinct sample sets. While the Halton sequence may generally be less accurate than the Sobol sequence, it exhibits less dependency on sample size $N$ compared to Sobol sequences, which require $N$ to be a power of 2~\cite{sobol1967distribution}. Owen’s randomization is well-suited for applications like TRRM because it allows for the efficient generation of new samples in parallel by constructing rows in the range $\left[N, N^\prime\right]$. Similar to Monte Carlo methods, the reproducibility of samples is ensured by seeding the local pseudo-random number generator used in the randomization process. Finally, Owen's randomization method was chosen for its straightforward implementation and lack of any additional user input.

This study builds upon recent work which implemented TRRM in OpenMC~\cite{scone_openmc_random_ray}, by also directly implementing the randomized Halton algorithm in OpenMC for more efficient sampling and direct integration with OpenMC's pseudo-random number generator. Owen's method, returns samples in $[0,1]$ in an array of size $N\times D$. Each dimension is constructed from a prime number which forms the basis of the sequence. In TRRM, we need to sample at-most five dimensions: three in space and two in angle. Therefore, only five primes need to be stored. Each iteration in TRRM will take samples from a separate randomized sequence which is specified with a unique ``batch seed''. This sets the seed for OpenMC's native pseudo-random number generator which is used to permute the sequence. Because Owen's method can generate $N^\prime$ new samples very efficiently by only generating the $[N,N^\prime]$ rows, the sampling algorithm is used to generate the subsequent row of the sequence each time a ray is emitted. The $1\times5$ array of samples in $[0,1]$ are then mapped to the ray's initial position and direction of travel.

% \begin{minipage}{0.7\textwidth}
%    \begin{algorithm}[H]
%     \caption{Randomized Halton (N, D, seed, skip=0)~\cite{Owen2017}} 
%     \label{alg:rhalton}
%     \begin{algorithmic}
%         \STATE Set seed
%         \STATE $\text{primes}=\left[2,3,5,7,11,13\right]$
%         \STATE Initialize rHalton to be an empty matrix of size $N$ by $D$
%         \FOR{d \textbf{in} [1,\dots, D]}
%             \STATE $b = \text{primes}[d]$
%             \STATE $\text{ind} = \left[\text{skip}, \text{skip}+1, \dots, \text{skip}+N\right]$
%             \STATE $\text{b2r}=1/b$
%             \STATE $\text{ans}=\text{ind}*0$
%             \STATE $\text{res}=\text{ind}$
%             \WHILE{$1-\text{b2r} < 1$}
%                 \STATE $\text{dig} = \text{res} \% b$
%                 \STATE $\text{pdig}=\text{random permutation}[\text{dig}]$
%                 \STATE $\text{ans} = \text{ans} + \text{pdig}*\text{b2r}$
%                 \STATE $\text{b2r}=\text{b2r}/b$
%                 \STATE $\text{res}=(\text{res}-\text{dig})/b$
%             \ENDWHILE
%             \STATE $\text{rHalton}[:, D]= \text{ans}$ 
%         \ENDFOR                    
%     \end{algorithmic} 
%    \end{algorithm}
% \end{minipage}

%%%%%%%%%%%%%%%%%%%%%%%%%%%%%%%%%%%%%%%%%%%%%%%%%%%%%%
\section{2D C5G7 RESULTS}\label{sec:results}
To evaluate the numerical and runtime performance of randomized Halton samples in TRRM, this study presents results from the 2D C5G7 k-eigenvalue benchmark, a seven-group quarter-core reactor problem~\cite{Lewis2001}. The benchmark consists of four assemblies, each made up of a $17\times17$ pin cell array of various fuels and control rods. Accuracy is measured by comparing the reference spatial pin power results to tally estimates using pseudo-random samples and randomized Halton samples. Results were generated on one CPU node (2x AMD EPYC 7763) utilizing one MPI rank and 256 OpenMP threads.

\begin{figure}[ht!]
    \centering
    \includegraphics[width=\textwidth]{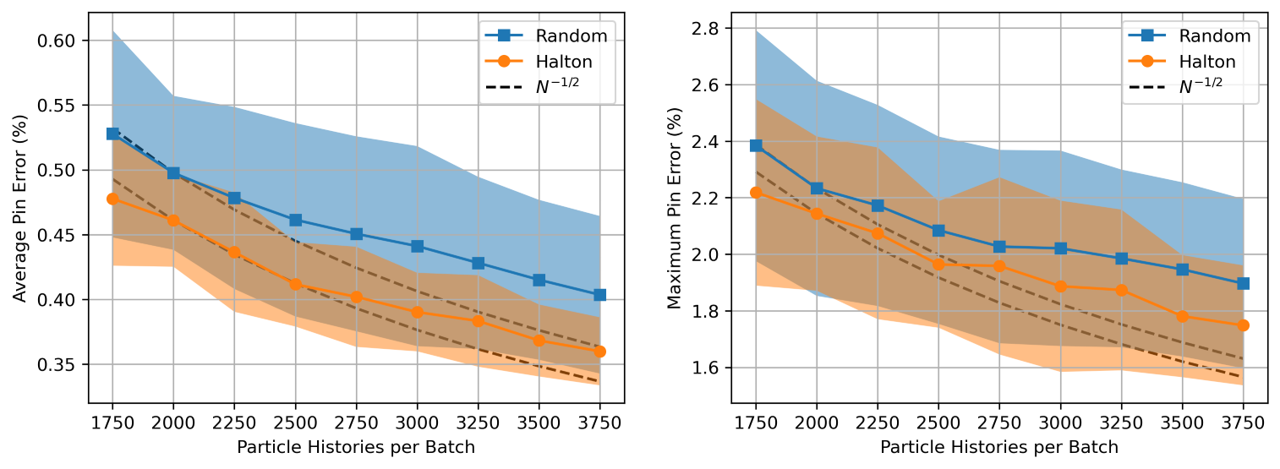}
    \caption{Average and maximum pin power error from an ensemble of 80 simulations as a function of particle histories per batch. The shaded areas represent one standard deviation of the ensemble results.}
    \label{fig:error}
\end{figure}

\begin{figure}[ht!]
    \centering
    \includegraphics[width=0.6\textwidth]{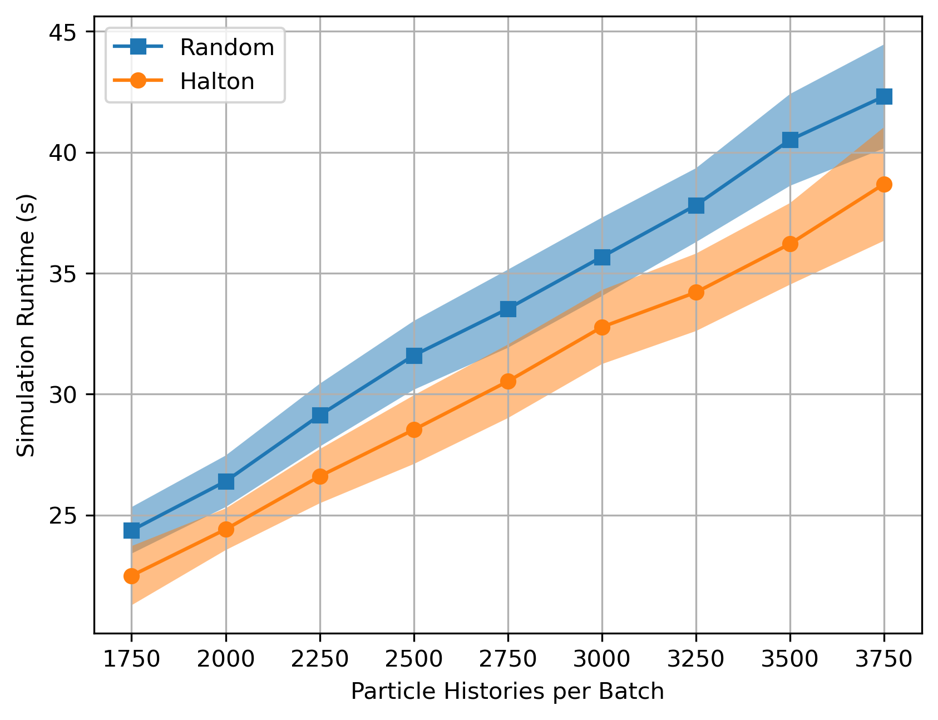}
    \caption{Mean simulation runtime from an ensemble of 80 simulations as a function of the number of particle histories per batch. The shaded area represents one standard deviation away from the mean of the ensemble.}
    \label{fig:runtime}
\end{figure}

The first numerical experiment presented compares the performance of the sampling methods with varying particle histories (rays) per batch. Each simulation used a dead length of 20 cm, a termination length of 200 cm, 600 inactive batches and 400 active batches. Additionally, each simulation utilized a flat-source approximation with 142,964 discretized regions. Finally, to better identify trends in the results, each data point represents the average value from an ensemble of 80 simulations, each with a unique starting seed. Figure~\ref{fig:error} shows the average and maximum pin power errors as a function of particle histories per batch. It is observed that the randomized Halton samples are consistently more accurate than the pseudo-random samples and with a lower standard deviation of the results. In addition to generating more accurate solutions, the randomized Halton samples also held a consistently faster runtime, shown in Figure~\ref{fig:runtime}.  This was an unexpected result and is further investigated in the second numerical experiment below. 

\begin{figure}[ht!]
    \centering
    \includegraphics[width=0.6\textwidth]{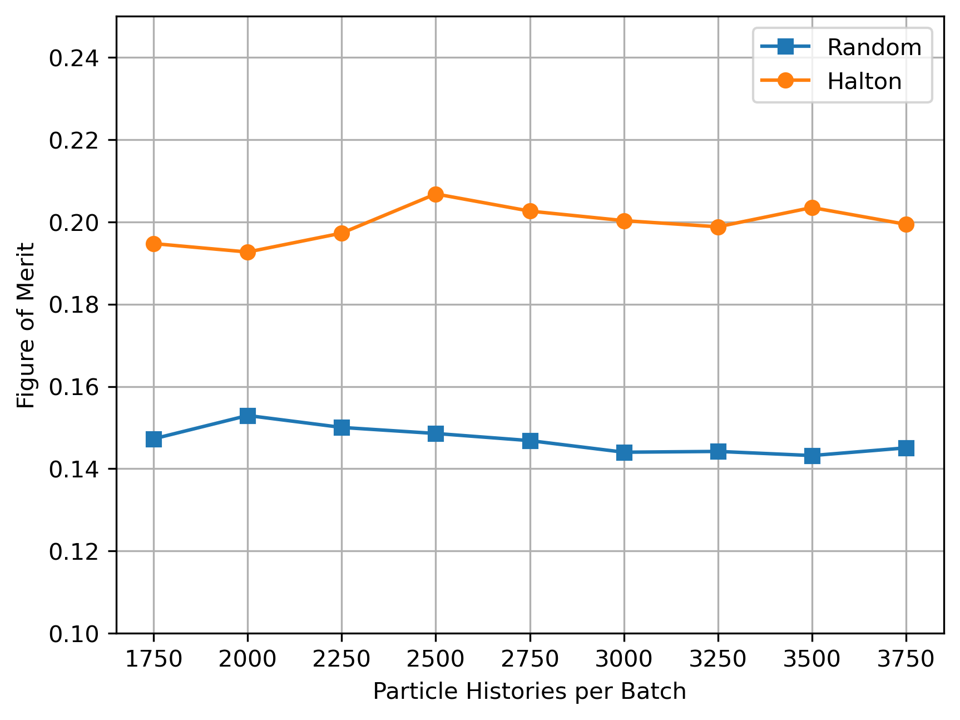}
    \caption{Average pin error figure of merit from an ensemble of 80 simulations as a function of particle histories per batch.}
    \label{fig:fom}
\end{figure}

The average pin power and simulation runtimes were combined in a figure of merit (FoM) defined as:
\begin{equation}
    \text{FoM}=\frac{1}{\epsilon^2 t}.
\end{equation}
Here, $\epsilon$ is the average pin power error and $t$ is the simulation runtime. Figure~\ref{fig:fom} shows a clear increase in the FoM when using randomized Halton samples. Notably however, the randomized Halton samples do not converge at the theoretical $O(N^{-1})$ expected from ideal QMC sampling. This could be partially due to the piecewise constant source approximation on a relatively coarse mesh. It could also be that some of the benefits of sampling emission sites with QMC are reduced when particles traverse many mean free paths.

\begin{figure}[ht!]
    \centering
    \includegraphics[width=0.6\textwidth]{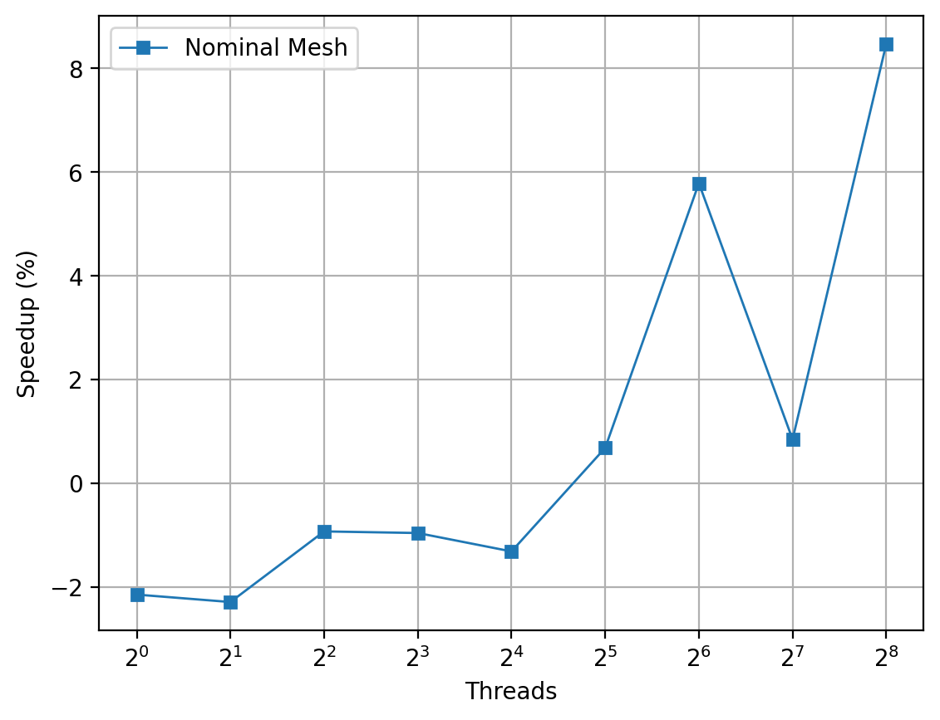}
    \caption{Percent speedup of the simulation runtime using Halton rays relative to random rays as a function of the the OpenMP thread count.}
    \label{fig:speedup}
\end{figure}

The native pseudo-random number generator in OpenMC is very computationally efficient~\cite{oneill2014pcg} and it is unlikely, although yet untested, that the randomized Halton sequence could generate samples more quickly. Instead, it is hypothesized that the decrease in runtime arises from reductions in shared memory contention when using Halton samples. Halton samples may provide increased shared memory efficiency due to the order in which points are sampled. For example, an ordinary 1D Halton Sequence in base three is given by $1/3$, $2/3$, $1/9$, $4/9$, $7/9$, $\dots$. Practically, this means that it is less likely that multiple threads will need access to the same memory at the same time. This would also imply a performance dependence on the number of OpenMP threads when using the randomized Halton sampling. To test this theory, a second numerical experiment was constructed to measure the performance as a function of thread count using the same 2D C5G7 construction from the first experiment. However, for practical runtime considerations, an ensemble of 80 simulations was run with 2000 particles per batch for 400 total batches. Results are shown in Figure~\ref{fig:speedup} which shows the percent speedup of the Halton samples relative to the random samples. When running serially, the Halton samples were approximately $2\%$ slower than the random samples. However, the percent speedup increases with thread count, crossing $0\%$ with 32 threads and ending with a $8\%$ \textit{speedup} at 256 threads.

%%%%%%%%%%%%%%%%%%%%%%%%%%%%%%%%%%%%%%%%%%%%%%%%%%%%%%
\section{CONCLUSIONS}\label{sec:conclusions}
In this study we have presented results from randomized Halton sampling, a form of randomized Quasi Monte Carlo sampling, implemented in The Random Ray Method in OpenMC. The Halton samples consistently produced more accurate solutions (over 10\% on average) in the 2D C5G7 problem than typical pseudo-random samples. When running simulations serially it appears that the Halton samples result in an approximate $2\%$ slowdown in runtime. However, as more shared memory threads are used, the Halton samples result in significant runtime speedup (observed up to $8\%$), owing to decreased shared memory contention. 

(R)QMC techniques are not often been employed in neutron transport simulations as the semi-deterministic nature of samples breaks the Markovian assumption needed for modeling the particle random walk. However, TRRM ray-trace procedure provides a well suited application for RQMC techniques. This work motivates further investigation in using RQMC techniques in TRRM. Additional work may include using the more accurate Sobol sequence in place of the Halton sequence or evaluating performance on highly sensitive IFBA pin-cell problems. TRRM's ``Immortal Ray'' variant may also benefit from RQMC sampling because it simulates rays for much shorter distances and therefore has an increased sensitivity to the starting location of ray.

%%%%%%%%%%%%%%%%%%%%%%%%%%%%%%%%%%%%%%%%%%%%%%%%%%%%%%
% \printnomenclature 
% If variables are extensively used in the text, a Nomenclature section would be helpful to the reader.
% \nomenclature{\(c\)}{Speed of light in a vacuum}
% \nomenclature{\(h\)}{Planck constant}
\section*{ACKNOWLEDGEMENTS}
This research used resources of the National Energy Research Scientific Computing Center, which is supported by the Office of Science of the U.S. Department of Energy under Contract No. DE-AC02-05CH11231.

\bibliographystyle{mc2025}
\bibliography{main}

% \appendix

% \section{}
% If necessary, include Appendices numbered in upper case alphabetical order.

% To ensure a uniform, professional look at the proceedings, please only modify the format of this template after checking with the publication chair first.

\end{document}